\documentstyle[11pt,IAU207_pasp,twoside,psfig]{article}
\def\edcomment#1{\iffalse\marginpar{\raggedright\sl#1\/}\else\relax\fi}
\reversemarginpar

\begin{document}

\title{Stellar Populations in Star Clusters: The R\^ole Played by Stochastic
Effects}
\author{Gustavo Bruzual A.}
\affil{Centro de Investigaciones de Astronom{\'\i}a (C.I.D.A.), Apartado Postal 264, M\'erida, Venezuela}

\begin{abstract}
In this paper I combine the results of a set of population synthesis
models with simple Montecarlo simulations of stochastic effects in the number of
stars occupying sparsely populated stellar evolutionary phases, to show that
the scatter observed in the magnitudes and colors of LMC and NGC 7252 star clusters
can be understood in the framework of current stellar evolution theory, without the need
to introduce ad-hoc corrections (e.g. artificially increasing the number of AGB stars). 
\end{abstract}

\section{Introduction}

When photometric properties of star clusters are compared to the predictions
of population synthesis models, it is common to find that the scatter among
the observed colors and the difference between these colors and the model
predictions, are larger than allowed by the observational errors.
This is particularly true for intermediate age clusters in optical-IR colors, like $V-K$,
but to a lesser extent the scatter is also large in $B-V$. 
Fig 1 illustrates this point.

In a recent paper, Maraston et al. (2001) show that the population synthesis models of
Maraston (1998) reproduce quite well the $V-K$ colors of the cluster population
observed in NGC 7252 at an age between 400 and 600 Myr. In these models, built on
the fuel consumption theorem (Renzini \& Buzzoni 1986), the contribution of AGB stars
to the bolometric light in the synthetic population has been increased over that expected
from stellar evolution theory, in such a way as to reproduce the empirical values
determined by Frogel et al. (1990) for a sample of LMC clusters as a function of SWB class.
Even though this is a valid procedure, in this paper I show that the optical and optical-IR
colors of the intermediate age clusters of the LMC and NGC 7252 are consistent with the
colors expected from models (Bruzual \& Charlot 2001; Bruzual 2000, 2001) based on current
stellar evolution theory, if stochastic fluctuations in the number of stars in
sparsely populated evolutionary stages are properly included into the models.
In what follows I use the Montecarlo technique pioneered by Barbaro \& Bertelli (1977),
Chiosi et al. (1988), Girardi et al. (1995), Santos \& Frogel (1997),
and Cervi\~no, Luridiana \& Castander (2000), to study the role that stochastic effects
on the initial mass function (IMF) have on star cluster colors.
I follow the Santos \& Frogel (1997) approach to study the LMC and NGC 7252 cluster samples.

\begin{figure} 
\centerline{\vbox{\psfig{figure=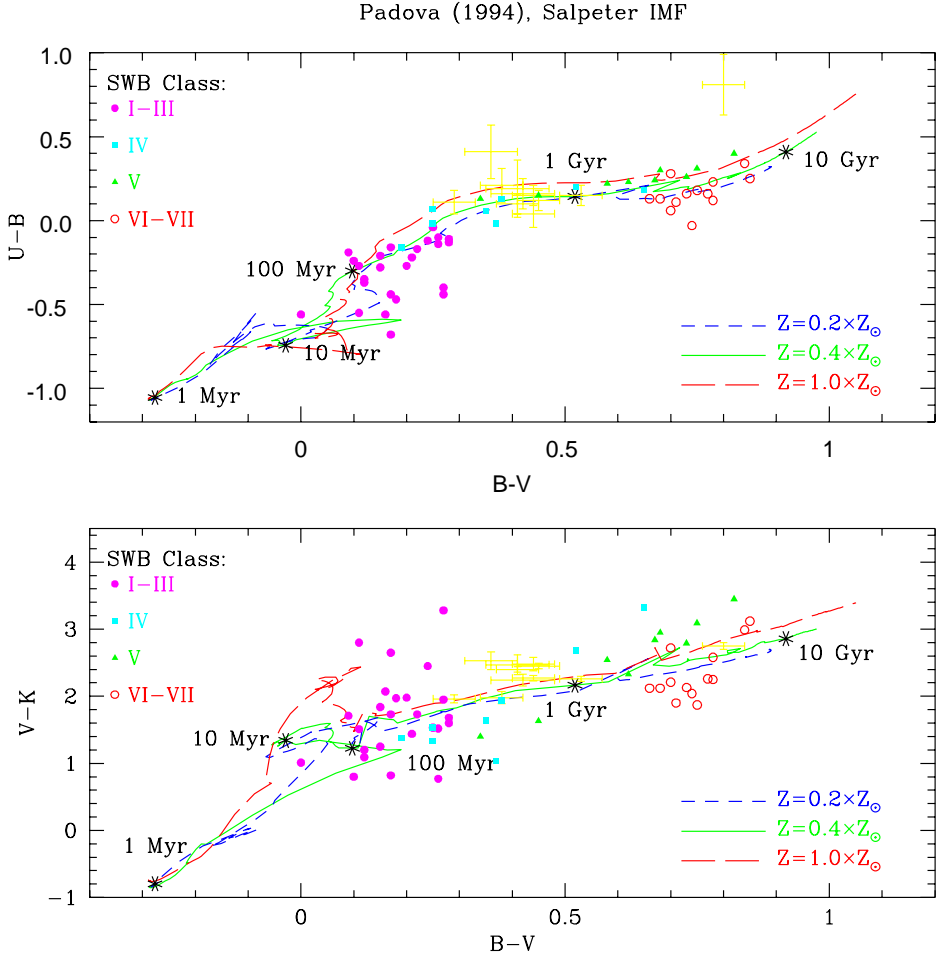,width=13.4cm}}}
\caption{
$U-B$ vs. $B-V$ and $V-K$ vs. $B-V$ color diagrams.
The solid dots, squares, triangles, and open circles represent
LMC globular clusters discriminated according to their SWB class, as
indicated in the upper left corner of each frame.
$B-V$ comes from van den Bergh (1981) and $V-K$ from Persson et al. (1983).
The points with error bars correspond to the star clusters in
NGC 7252 from Maraston et al. (2001).
The lines represent the evolution in this plane of
BC2000 SSP models (Bruzual 2000) computed using the Padova (Fagotto et al. 1994a,b,c)
evolutionary tracks
for $Z = 0.2\times Z_\odot, 0.4\times Z_\odot$, and $Z_\odot$, the Salpeter (1955) IMF,
and the Lejeune et al. (1998) stellar atlas.
The * symbols along the $Z = 0.4\times Z_\odot$ line mark the model colors at the age indicated
by the labels, and can be used to roughly date the clusters.
}
\end{figure}

\section{Method}

The derivation below follows closely Santos \& Frogel (1997). The IMF
\begin{equation}
\Phi(m) = dN/dm = Cm^{-(1+x)},
\end{equation}
normalized as usual,
\begin{equation}
C = { x \over {m_l^{-x} - m_u^{-x} } },
\end{equation}
obeys, $\Phi(m) \ge 0$, and
\begin{equation}
\int_{m_l}^{m_u}\Phi(m')dm' = 1.
\end{equation}
$\Phi(m)$ can be interpreted as a probability distribution function which gives the probability
that a random mass $m'$ is in the range between $m$ and $m+dm$.
$\Phi(m)$ can be transformed into another probability distribution function $g(N)$, such that
the probability of occurrence of the random variable
$N'$ within $dN$ and the probability of occurrence of the random variable $m'$ within $dm$
is the same,
\begin{equation}
|\Phi(m)dm| = |g(N)dN|.
\end{equation}
where $N$ is a single-valued function of $m$. From (1),
\begin{equation}
N(m) = \int_{m_l}^{m}\Phi(m')dm',
\end{equation}
is a cumulative distribution function which 
gives the probability that the mass $m'$ is less or equal to $m$, and using (4),
it follows that
\begin{equation}
g(N) = 1, ~~~~ 0 \le N \le 1.
\end{equation}
$g(N)$ is thus a uniform distribution for which any value is equally likely in
the interval $ 0 \le N \le 1$. If we sample $N$ using a random number generator
(Press et al. 1992), from (5) we can obtain $m$ as a function of $N$,
\begin{equation}
m = [(1-N)m_l^{-x} + Nm_u^{-x}]^{- {1 \over x} }.
\end{equation}
For each star of mass $m$ generated in this way, we obtain its observational properties
from the log $T_{eff}$ and log $L$ corresponding to this mass in the isochrone at the age
of interest.
We repeat the procedure until the cluster mass, i.e. the sum of $m$ for all the stars generated,
including dead stars, reaches the desired value.
Adding the contribution of each star to the flux in different photometric bands, we obtain the
cluster magnitude and colors in these bands.
Throughout this paper I use $x=1.35$ (Salpeter 1955), and for the lower and upper mass limits
of star formation, I assume $m_l=0.09$ and $m_u=125$ M$_\odot$, respectively. 

\begin{figure} 
\centerline{\vbox{\psfig{figure=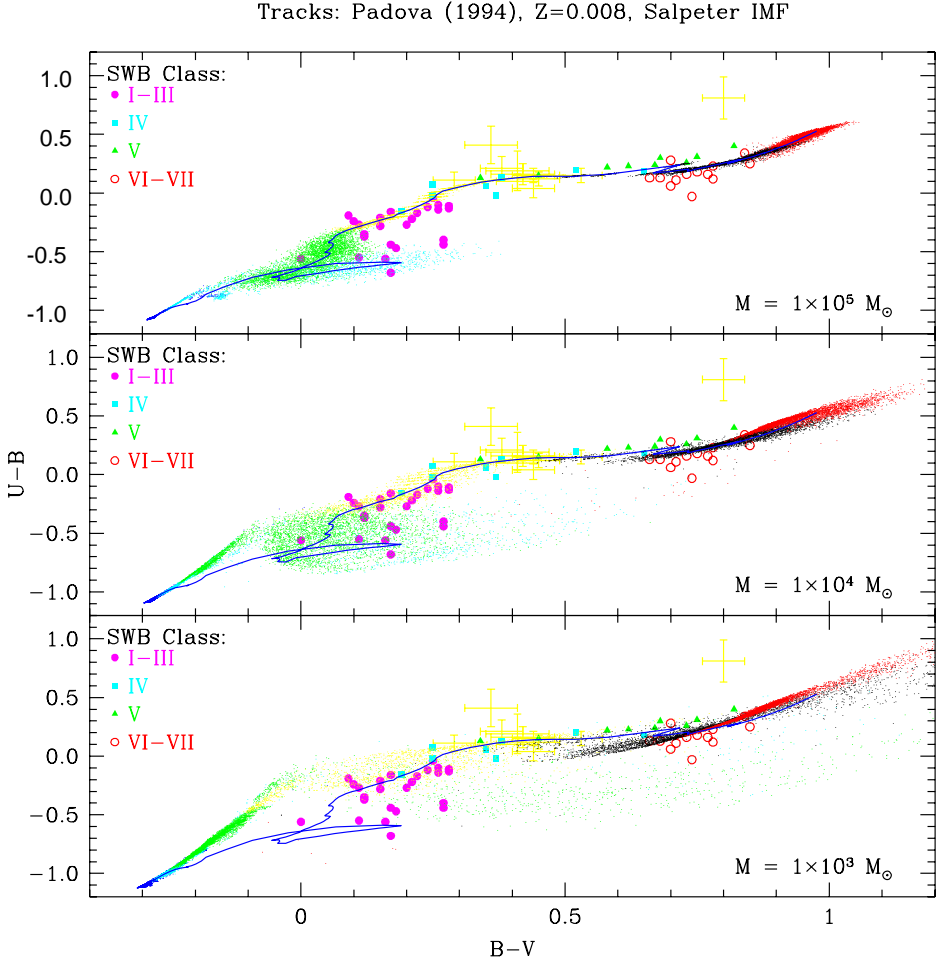,width=13.4cm}}}
\caption{
The small dots indicate the $U-B$ and $B-V$ colors resulting from
different simulations in which stochastic fluctuations in the number of
stars that populate different evolutionary stages are taken into account
as indicated in the text. 100 simulations were run at each of 220 time
steps or isochrones, obtained from the BC2000 SSP model (Bruzual 2000) for the Padova (Fagotto et al. 1994a,b,c)
tracks for $Z = 0.4\times Z_\odot$, the Salpeter (1955) IMF, and the Lejeune et al.
(1998) stellar atlas (uncorrected atmospheres).
The simulation is stopped when the total cluster mass (including
dead stars) reaches $1\times10^5,\ 1\times10^4,$ and $1\times10^3$ M$_\odot$
(top to bottom), as indicated in the lower right corner of each frame.
The fluctuations in the colors become larger as the cluster mass
decreases.
The number of stars in the $1\times10^3$ M$_\odot$ cluster may be
unrealistically small, and some evolutionary stages are not sampled.
The data points and the solid line for $Z = 0.4\times Z_\odot$ are the same as in Fig 1.
The solid line represents the evolution of this model with no fluctuations,
which is equivalent to an infinite number of stars populating the IMF.
}
\end{figure}

\begin{figure} 
\centerline{\vbox{\psfig{figure=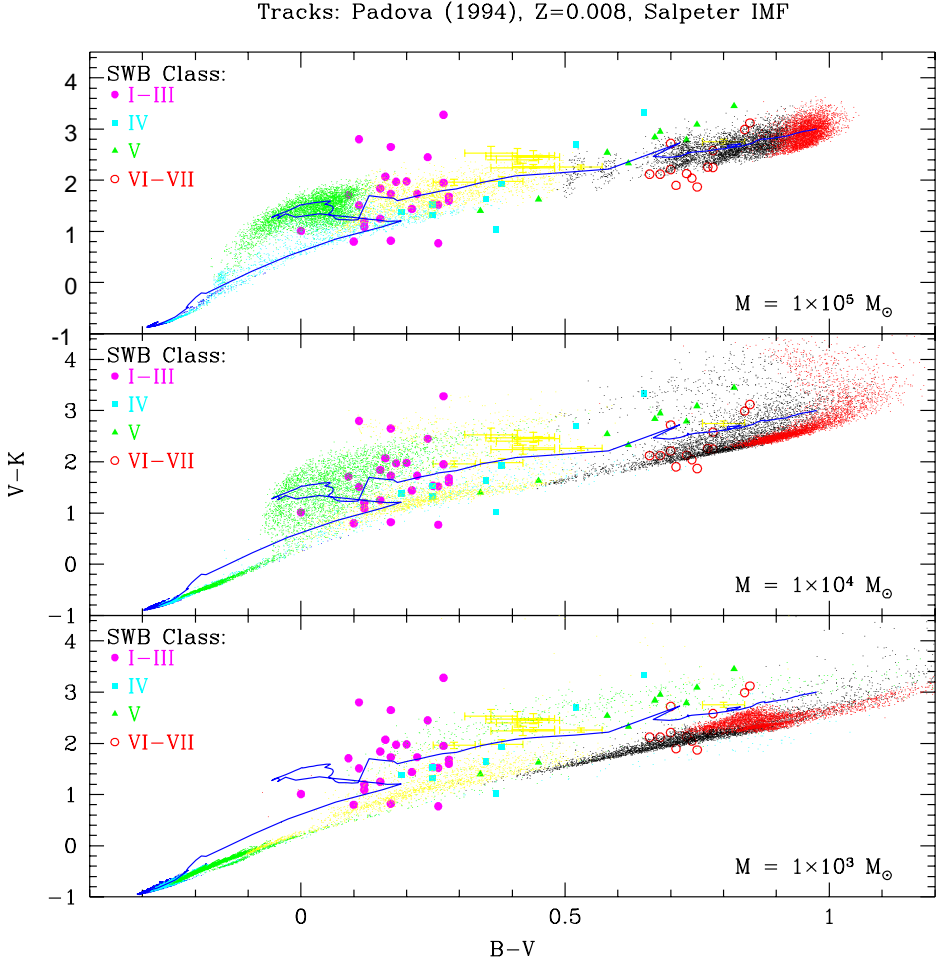,width=13.4cm}}}
\caption{
The small dots indicate the $V-K$ and $B-V$ colors resulting from
different simulations in which stochastic fluctuations in the number of
stars that populate different evolutionary stages are taken into account
as indicated in the text. 100 simulations were run at each of 220 time
steps or isochrones, obtained from the BC2000 SSP model (Bruzual 2000) for the Padova (Fagotto et al. 1994a,b,c)
tracks for $Z = 0.4\times Z_\odot$, the Salpeter (1955) IMF, and the Lejeune et al.
(1998) stellar atlas (uncorrected atmospheres).
The simulation is stopped when the total cluster mass (including
dead stars) reaches $1\times10^5,\ 1\times10^4,$ and $1\times10^3$ M$_\odot$
(top to bottom), as indicated in the lower right corner of each frame.
The fluctuations in the colors become larger as the cluster mass
decreases.
The number of stars in the $1\times10^3$ M$_\odot$ cluster may be
unrealistically small, and some evolutionary stages are not sampled.
The data points and the solid line for $Z = 0.4\times Z_\odot$ are the same as in Fig 1.
The solid line represents the evolution of this model with no fluctuations,
which is equivalent to an infinite number of stars populating the IMF.
The expected fluctuations in $V-K$ for a $1\times10^4$ M$_\odot$ cluster
amount to almost 2 magnitudes, in close agreement with the range of
colors observed at a given age.
}
\end{figure}
\section{Results}

Figs 2 to 6 show the results of the Montecarlo simulations. For reasons of space the
details of each figure given in the caption are not repeated in the text. From
Figs 2 and 3 it is apparent that the fluctuations in the colors become larger as
the cluster mass decreases.
The number of stars in the $1\times10^3$ M$_\odot$ cluster may be
unrealistically small, and some evolutionary stages are not sampled.
There are not enough high mass MS stars in the $1\times10^3$ M$_\odot$ and
$1\times10^4$ M$_\odot$ clusters to make them as blue in $U-B$ at early ages as the model
computed with the analytical IMF (equivalent to infinite number of stars).
In the $1\times10^5$ M$_\odot$
case the upper MS is well sampled and both models are equally blue in $U-B$.
Fig 3 shows that the lower mass cluster models at early ages are bluer in $V-K$
than the analytic model because the lower MS is over populated with respect to the short
lived phases with redden this color. For the $1\times10^5$ M$_\odot$ cluster the
agreement is good at early ages.
The expected fluctuations in $V-K$ for a $1\times10^4$ M$_\odot$ cluster
amount to almost 2 magnitudes, in close agreement with the range of
colors observed at a given age, and is considerably broader than for the $1\times10^4$ M$_\odot$
case.
Fig 4 shows an enlargement of the diagrams for the $1\times10^4$ M$_\odot$ star cluster.
At the intermediate ages the models are redder in $V-K$ than the analytic IMF model
because of a larger number of AGB stars, which appear naturally as a consequence
of stochastic fluctuations in the IMF.
The regions in these planes occupied by the observations and
the cluster simulations overlap quite well.
This comparison seems to favor a not very large mass for these clusters, if stochastic
fluctuations are responsible of the color variations.
This conclusion is supported by Fig 5 in which I plot $K$ vs $V-K$ at the distance of the LMC.
The region covered by the $1\times10^4$ M$_\odot$ clusters matches the region covered by
the observations very well.
This is not the case for the $1\times10^5$ M$_\odot$ clusters, which show
a much lower dispersion in color for the same $K$ magnitude than the observations and the
lower mass simulations.

\begin{figure} 
\centerline{\vbox{\psfig{figure=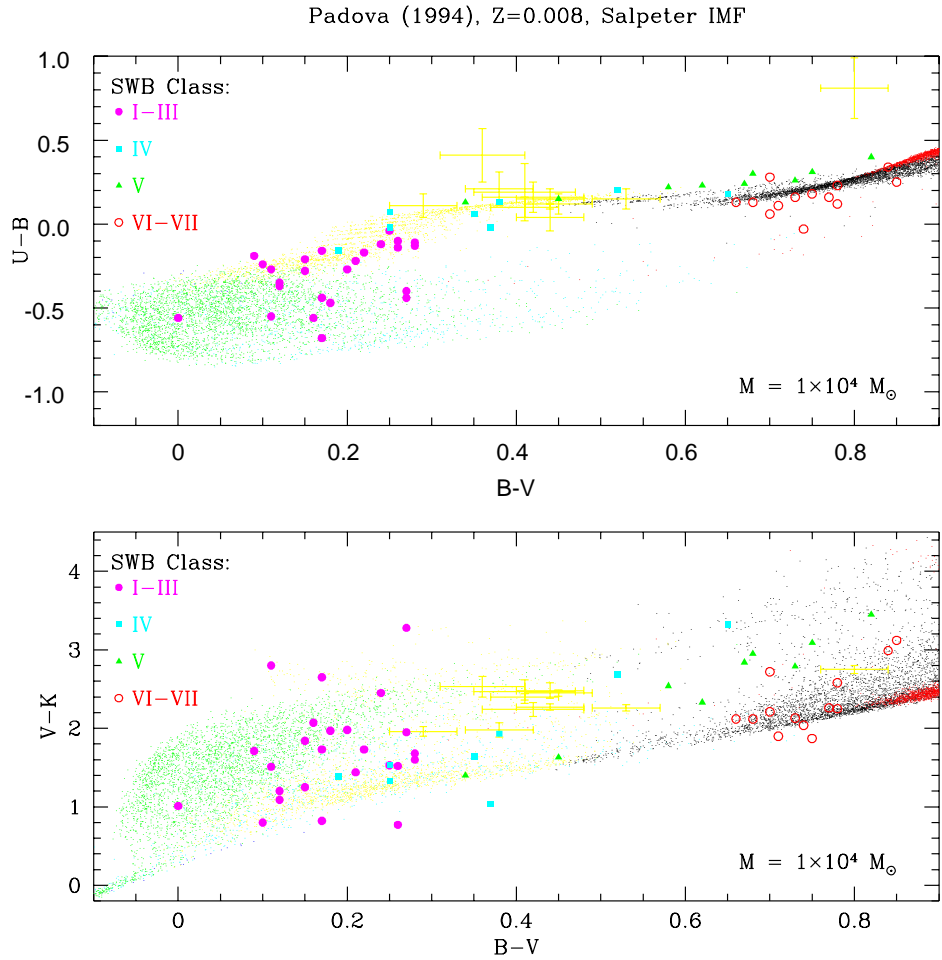,width=13.4cm}}}
\caption{
Enlarged view of the $U-B$ vs. $B-V$ and $V-K$ vs. $B-V$ color
diagrams shown in the middle frame of Figs 2 and 3.
The data points are the same as in Fig 1.
The regions in these planes occupied by the observations and
the cluster simulations overlap very well.
}
\end{figure}

\begin{figure} 
\centerline{\vbox{\psfig{figure=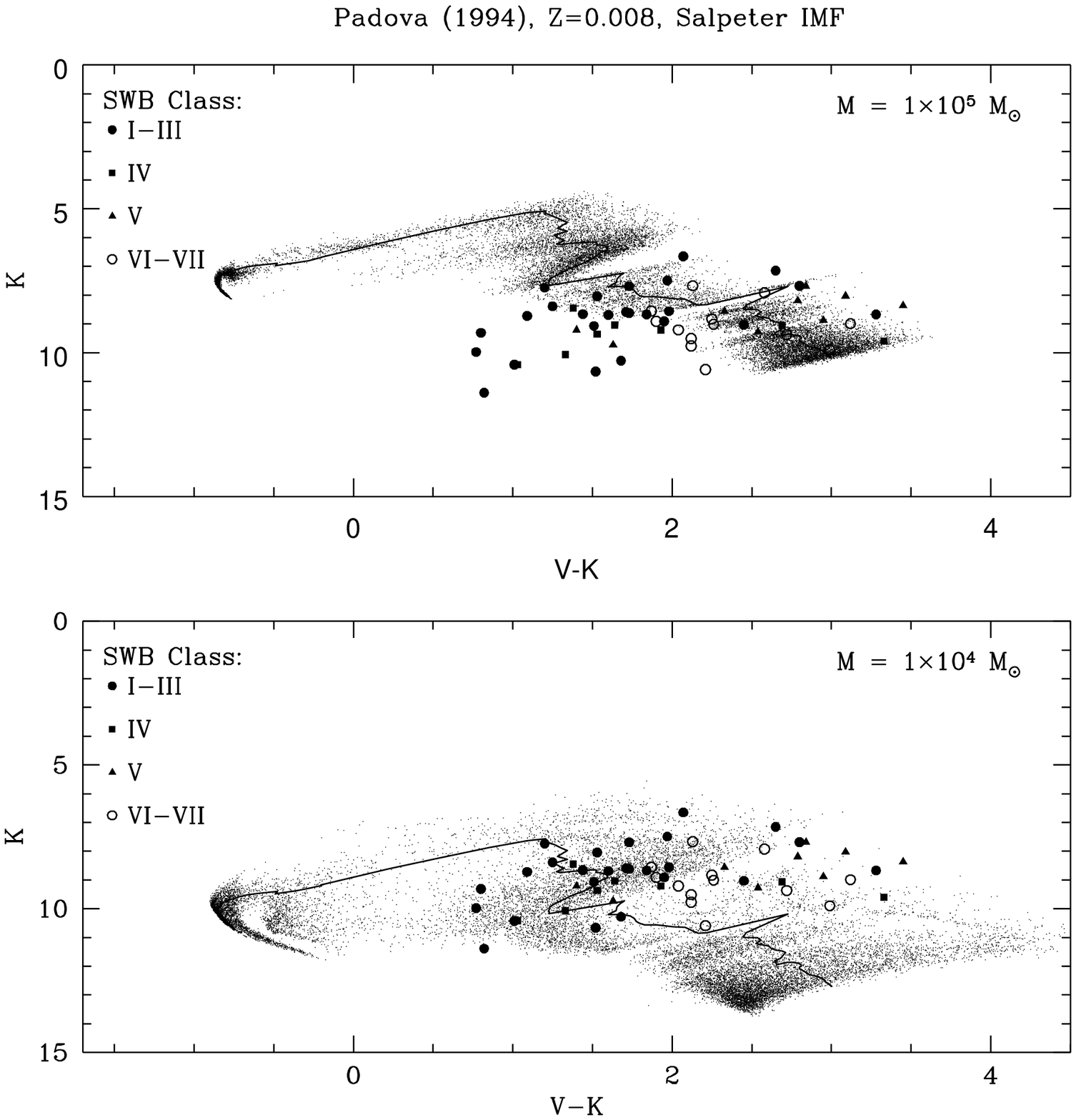,width=13.4cm}}}
\caption{
The small dots indicate the $K$ magnitude (at the distance of the LMC)
and $V-K$ color resulting from
different simulations in which stochastic fluctuations in the number of
stars that populate different evolutionary stages are taken into account
as indicated in the text. 100 simulations were run at each of 220 time
steps or isochrones, obtained from the BC2000 SSP model (Bruzual 2000) for the Padova
(Fagotto et al. 1994a,b,c) tracks for $Z = 0.4\times Z_\odot$, the Salpeter (1955) IMF,
and the Lejeune et al.  (1998) stellar atlas (uncorrected atmospheres).
The simulation is stopped when the total cluster mass (including
dead stars) reaches $1\times10^5$ (top) and $1\times10^4$ M$_\odot$ (bottom),
as indicated in the upper right corner of each frame.
The fluctuations in the colors get considerably larger as the cluster mass
decreases.
The data points are the same as in Fig 1.
The solid line represents the evolution of this model with no fluctuations,
which is equivalent to an infinite number of stars populating the IMF.
}
\end{figure}

\section{Conclusions}

Even though the Montecarlo technique used in this paper has been used successfully
before to explain cluster colors, I have revisited it in this paper to show that
there is no need to introduce ad-hoc assumptions into population synthesis models
(Maraston 1998, Maraston et a. 2001), which represent a departure from our current
understanding of stellar evolution theory, in order to explain the observed
range of values of cluster colors and magnitudes.
In this paper I have briefly summarized the results of simple simulations
which show clearly how the range of colors observed in intermediate age star clusters
can be understood on the basis of current stellar evolution
theory, if we take into account properly the expected variation in the number
of stars occupying sparsely populated evolutionary stages due to stochastic
fluctuations in the IMF.

Our preliminary conclusion that cluster masses in the range around $1\times10^4$ M$_\odot$
are preferred, as well as the dependence on this mass of the fluctuations in magnitude and
colors will be explored in detail by Bruzual \& Charlot (2001).

The interested reader may request results for other quantities equally sensitive to stochastic
fluctuations, like the number of ionizing photons from a young stellar population, not shown here
for lack of space (see also Cervi\~no, Luridiana \& Castander 2000).

\end{document}